# Crucible aperture: an effective way to reduce source oxidation in oxide molecular beam epitaxy process

Yong-Seung Kim[1], Namrata Bansal[2], and Seongshik Oh[1, a]

[1] Department of Physics & Astronomy, Rutgers, The State University of New Jersey, 136 Frelinghuysen Rd, Piscataway, NJ, U.S.A.

[2] Department of Electrical and Computer Engineering, Rutgers, The State University of New Jersey, 94 Brett Rd, Piscataway, NJ, USA

a) Electronic mail: ohsean@physics.rutgers.edu

MATERIAL NAMES: Strontium (Sr), Strontium Oxide (SrO), Oxygen ($O_2$)

# Abstract

Growing multi-elemental complex-oxide structures using an MBE (Molecular Beam Epitaxy) technique requires precise control of each source flux. However, when the component elements have significantly different oxygen affinities, maintaining stable fluxes for easily oxidizing elements is challenging because of a source oxidation problem. Here, using Sr as a test source, we show that a crucible aperture insert scheme significantly reduces the source oxidation in an oxide-MBE environment. The crucible aperture insert was shaped like a disk with a hole at the center and was mounted inside the crucible; it blocks most of the oxygen species coming to the source, thus reducing the source oxidation. However, the depth of the aperture disk was critical for its performance; an ill-positioned aperture could make the flux stability even worse. With an optimally positioned aperture insert, the crucible exhibited more than four times improvement in Sr flux stability, compared to a conventional, non-apertured crucible.

## I. INTRODUCTION

Complex-oxide MBE process uses multiple source materials having significantly different oxygen affinities[1]. But source oxidation problem leads to flux instability[2], and maintaining stable fluxes in an oxygen environment for all elements is a challenging task[3-4]. In case of Ca and Sr, when exposed to strong oxidation condition, they tend to oxidize so much[2] that a real-time flux monitoring scheme such as atomic absorption spectroscopy (AA)[5-6] is required to control their fluxes better than 1%. As the number of component elements grows, such a real-time monitoring scheme becomes cumbersome to implement, and makes the complex-oxide growth very complicated. Ideally if all the fluxes can be made completely stable through the entire growth, then no real-time monitoring will be needed and the complex-oxide MBE process will become a much simpler practice.

In our previous work, we studied the effect of source geometry, flux rate, and oxygen pressure on flux stability[7]. It showed that extended port geometry compared to a standard port, higher flux compared to lower flux, and un-melted source shape compared to melted-source enhanced the flux stability. Based on our understanding of the source oxidation problem from this previous study, we devised a simple crucible-aperture scheme as a way to minimize the source oxidation. The aperture insert is a disk having a hole at the center. It minimizes the source area that is exposed to the oxygen species and blocks out most of those coming directly to the source. So far, various kinds of crucible inserts have been proposed for different purposes in conventional semiconductor MBE systems. Maki *et. al* proposed a conical insert to eliminate the flux transients[8], Thorpe *et. al* used a tilted conical insert to improve thickness uniformity of GaAs epitaxial layers[9], and Sacks *et. al* reported stacked disk inserts to reduce visible defect densities[10]. However, no such studies have been reported for an oxide MBE process, and here we show that the crucible aperture is a very effective way to minimize source oxidation in the oxide-MBE environment. This scheme is so simple that it can be easily implemented in any existing oxide-MBE system.

## II. EXPERIMENT

Our experiments were performed in a custom-designed SVTA MOS-V-2 MBE system. Sr sources were thermally evaporated from hot-lip low-temperature effusion cells (SVTA-275/450/458-XX). Stability of the source temperature was better than 0.1 °C, and flux drift was less than 1 % over several hours when no oxygen was introduced. Strontium (Aldrich-APL, 99.99 %) was loaded in a pyrolytic boron nitride (PBN) crucible and pre-melted to have a smooth top surface (service provided by Aldrich-APL). Oxygen partial pressure was controlled from $10^{-9}$ to $10^{-4}$ Torr using a differentially-pumped mass flow controller in combination with a precision leak valve. The base pressure of the system was ~$10^{-10}$ Torr during the experiments.

The Sr flux was measured using a water-cooled quartz crystal microbalance (QCM: Inficon BDS-250, XTC/3) mounted at the growth position. Because the QCM signal was strongly affected by heat load from the source during opening and closing of the shutter[7], we excluded the first one hour data for all the long-term stability analyses. In addition, five minute moving average was used for all long-term results to improve the resolution and signal-to-noise ratio[7].

We employed various types of apertures to study the effect of the crucible aperture on flux stability. Two different materials (alumina and tantalum) and three different aperture-diameters (1.5, 3.1 and 5.0 mm) were employed, alumina apertures were provided by SVTA and tantalum apertures were home-made. The thickness of alumina aperture was 1 mm and that of tantalum was 25 µm. The tantalum apertures had small cuts 2 mm apart and 5 mm deep around the circumference for flexibility and thus can be positioned at almost any depth in the crucible; see Figure 1. For alumina apertures, two alumina rings were used to fix the aperture insert at a desired position, and because of their rigidity, the mounting position of the alumina aperture was relatively limited compared to the tantalum one.

## III.  RESULTS AND DISCUSSION

Figure 2 shows the effect of the aperture depth and its size on short term flux stability. First, we note that at a fixed aperture depth of 3.5 cm, the source temperatures necessary to provide similar flux values strongly depend on the insert type. With the alumina insert, the source temperature had to be significantly raised to obtain a flux value comparable to the one without any aperture, and this can be easily explained by the reduced source area due to the small aperture. But with the tantalum insert, the flux remained almost unchanged from the value of a non-apertured crucible at the same source temperature. This implies that tantalum, being a metal, reflects most of the thermal radiation from the hot source back to the source surface and raises its effective temperature, resulting in enhanced flux, and this effect offsets the reduced source area.

Figure 2 also shows that with the same insert type, as the insert got closer to the crucible orifice, maintaining comparable fluxes required higher source temperatures; this is because the temperature of the insert drops near the orifice. Obviously, smaller aperture size also resulted in reduced flux. In terms of oxygen pressure dependence, Figure 2 shows that the aperture depth is the most critical factor -- the deeper the aperture is, the more sensitive the flux is to the oxygen pressure. As reported in our previous study, as the insert depth increases, the flux contribution from the crucible wall as a secondary source also increases and this results in a more severe short-term source oxidation[7].

With respect to the source oxidation, figure 2 shows that the insert positioned at the crucible orifice gives the best performance. In this case, however, severe material build-up occurred underneath the aperture. Lower temperature of the crucible orifice than that of the source surface is the main cause for this problem. This material build-up choked the aperture and induced serious fluctuation and drift in the source flux as shown in Fig. 3. The fluctuation amplitude was about 2 % of the flux value and the drift was more than 8 % during three hour operation. Furthermore, the build-up required a significantly higher source temperature to maintain a similar flux value. But the increase in the source temperature worsens the situation because higher temperature accelerates the build-up process. After finishing many hours of long term stability test, we found that the entire Sr source in the crucible was condensed underneath the aperture insert. With our hot-lip cell, the aperture insert had to be positioned deeper than 3 cm to prevent this problem. But when it is too deep, the crucible wall above the aperture works as an easily-oxidizing secondary source as mentioned above[7]. Because the best source stability requires no material build-up and minimum source oxidation, 3.5 cm was chosen as the optimal depth of the aperture insert. With this optimal aperture configuration, we tested long-term flux stability over 3 hours as shown in Fig. 3. The crucible with an aperture insert (alumina, opening diameter = 5 mm, depth = 3.5 cm) exhibited significantly better flux stabilities than the one without an aperture. For example, in $10^{-5}$ Torr of molecular oxygen, while the crucible without an insert exhibited 5.5 % decrease in Sr flux at a growth rate of ~0.05 Å/sec, the one with an insert showed only 1.2 % reduction over the same period [Fig. 3]. This implies that the insert improved the flux stability more than four times. At a higher growth rate of ~0.13 Å/sec, which is still slower than a typical growth rate, the flux variation was less than 1 % over a three hour period. Considering that source oxidation becomes weaker as the growth rate increases [7], the aperture scheme will provide flux stabilities much better than 1 % at typical growth rates.

## IV. CONCLUSIONS

An aperture scheme has been proposed and evaluated in the context of source oxidation problem in oxide-MBE processes. With an optimal aperture depth, the flux stability was significantly enhanced compared to that of a non-apertured crucible. We achieved a stable Sr flux (less than 1 % flux variation) over three hours even at a slower-than-typical growth rate in an oxygen pressure of $1\times10^{-5}$ Torr. This scheme will provide one of the easiest and cheapest solutions for the source oxidation problem in an oxide MBE environment.

# ACKNOWLEDGMENTS

This work is supported by IAMDN of Rutgers University, National Science Foundation (NSF DMR-0845464) and Office of Naval Research (ONR N000140910749).

# Figure Captions

**Fig. 1.** Schematic diagram of the alumina and tantalum aperture inserts. While the alumina insert was mounted with two rings, the tantalum insert was fixed into a position by its elastic force.

**Fig. 2.** Dependence of the short-term stability on the aperture depth at comparable flux values. Source oxidation occurred more significantly as the insert depth increased. But when the insert was mounted near the crucible orifice, material build-up occurred underneath the aperture as shown below in Fig. 3.

**Fig. 3.** Long-term flux stability for crucibles with and without an aperture insert. The crucible with an aperture insert, positioned at an optimal depth, exhibited significantly better stability than the one without an insert. But when the insert was positioned too close to the orifice, the flux became unstable as the top-most curve shows.

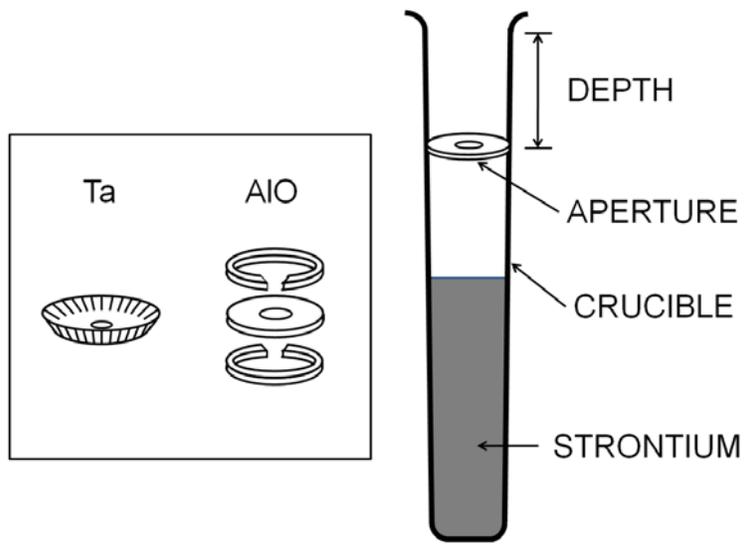

**Fig. 1**

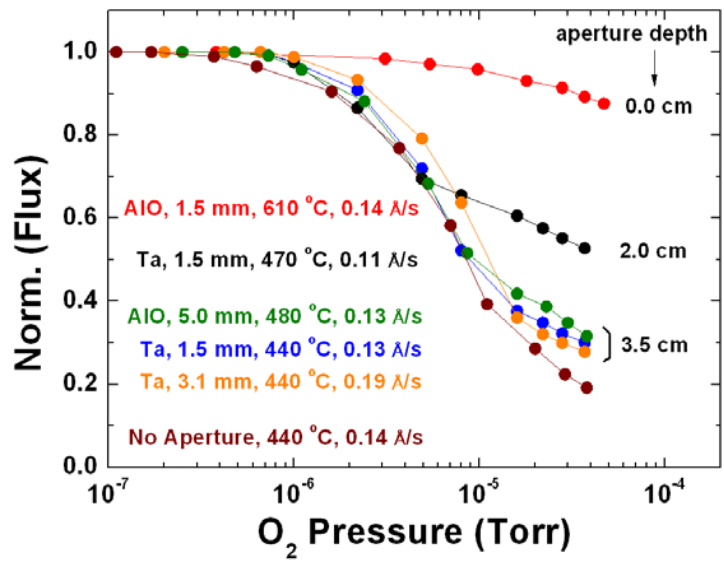

**Fig. 2**

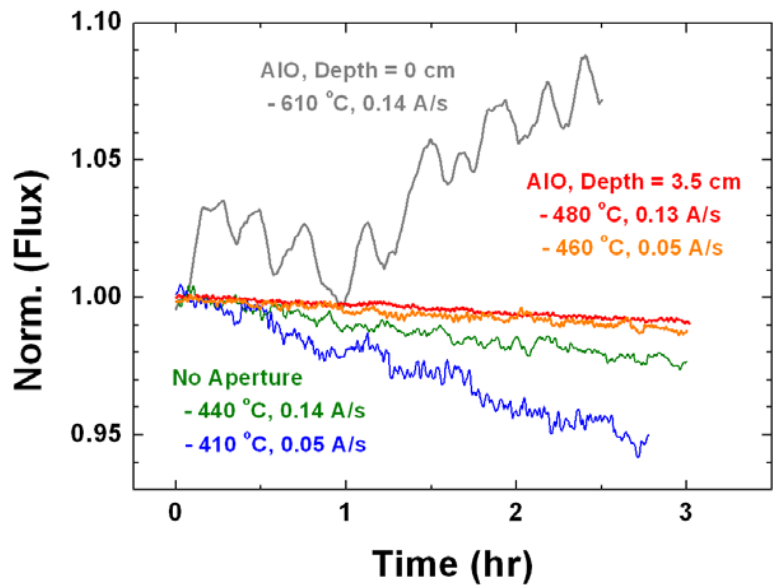

Fig. 3.

# References

[1] S. Oh, M. Warusawithana, and J. N. Eckstein, Physical Review B **70**, 064509 (2004).

[2] E. S. Hellman and E. H. Hartford, Journal of Vacuum Science & Technology B **12**, 1178 (1994).

[3] J. N. Eckstein and I. Bozovic, Annual Review of Materials Science **25**, 679 (1995).

[4] D. J. Rogers, P. Bove, and F. H. Teherani, Superconductor Science & Technology **12**, R75 (1999).

[5] M. E. Klausmeierbrown, J. N. Eckstein, I. Bozovic, and G. F. Virshup, Applied Physics Letters **60**, 657 (1992).

[6] Y. Kasai and S. Sakai, Review of Scientific Instruments **68**, 2850 (1997).

[7] Y. S. Kim, N. Bansal, C. Chaparro, H. Gross, and S. Oh, Journal of Vacuum Science & Technology A **28**, 271 (2010).

[8] P. A. Maki, S. C. Palmateer, A. R. Calawa, and B. R. Lee, Journal of Vacuum Science & Technology B **4**, 564 (1986).

[9] A. J. S. Thorpe, A. Majeed, C. J. Miner, Z. R. Wasilewski, and G. C. Aers, Journal of Vacuum Science & Technology a-Vacuum Surfaces and Films **9**, 3175 (1991).

[10] R. N. Sacks, G. A. Patterson, and K. A. Stair, Journal of Vacuum Science & Technology B **14**, 2187 (1996).